\documentclass{article}

\usepackage{microtype}
\usepackage{graphicx}
\usepackage{subcaption}
\usepackage{booktabs} 

\usepackage{stfloats}
\usepackage{enumitem}
\usepackage{xspace}
\usepackage{graphbox}
\usepackage{multirow}
\usepackage{tablefootnote}

\usepackage{hyperref}

\newcommand{\kbenchsyz}{\textsc{kBenchSyz}\xspace}

\newcommand{\sysplus}{\textsc{kGym++}\xspace}
\newcommand{\sys}{\textsc{kGym}\xspace}
\newcommand{\agent}{\textsc{kAgent}\xspace}
\newcommand{\data}{\textsc{kBenchSyz}\xspace}

\newcommand{\cache}{\textsc{kCache}\xspace}

\newcommand{\tprintk}{\texttt{trace\_printk}\xspace}
\newcommand{\debugtree}{$\mathbb{T}_{debug }$\xspace}
\newcommand{\syzkaller}{Syzkaller\xspace}

\newcommand*\circled[1]{\tikz[baseline=(char.base)]{
            \node[shape=circle,draw,inner sep=0.75pt] (char) {#1};}}

\usepackage{relsize}

\usepackage[accepted]{icml2026}

\usepackage{pifont, xcolor}
\usepackage{amsmath}
\usepackage{amssymb}
\usepackage{mathtools}
\usepackage{amsthm}
\usepackage[dvipsnames]{xcolor}

\usepackage{xspace}
\usepackage{enumitem}
\usepackage{pifont}
\usepackage{hhline}
\usepackage{boldline}
\usepackage{tikz}
\usepackage{listings}
\usepackage{mdframed}
\usepackage[most]{tcolorbox}
\usepackage{titlesec}
\usepackage[belowskip=-5pt,aboveskip=0pt]{caption}

\definecolor{lightmauve}{rgb}{0.86, 0.82, 1.0}

\usepackage{lineno} 
\makeatletter
\@namedef{ver@lineno.sty}{9999/12/31}
\@namedef{opt@lineno.sty}{}
\makeatother
\usepackage[]{minted} 

\usepackage[capitalize,noabbrev]{cleveref}

\theoremstyle{plain}

\theoremstyle{definition}

\theoremstyle{remark}

\usepackage[textsize=tiny]{todonotes}

\icmltitlerunning{\agent: An execution-guided crash resolution agent for the Linux kernel}

\begin{document}

\twocolumn[
  \icmltitle{\agent: An execution-guided crash resolution agent for the Linux kernel}



  \icmlsetsymbol{equal}{*}
  \icmlsetsymbol{equaltwo}{\#}

  \begin{icmlauthorlist}
    \icmlauthor{Alex Mathai}{equal,cu}
    \icmlauthor{Chenxi Huang}{equal,cu}
    \icmlauthor{Suwei Ma}{equaltwo,cu}
    \icmlauthor{Jihwan Kim}{equaltwo,cu}
    \icmlauthor{Hailie Mitchell}{cu}
    \icmlauthor{Aleksandr Nogikh}{googlede}
    \icmlauthor{Petros Maniatis}{gdm}
    \icmlauthor{Franjo Ivan\v{c}i\'{c}}{googlenj}
    \icmlauthor{Junfeng Yang}{cu}
    \icmlauthor{Baishakhi Ray}{cu}
  \end{icmlauthorlist}

  \icmlaffiliation{cu}{Department of Computer Science, Columbia University, New York, NY, USA}
  \icmlaffiliation{googlenj}{Google Inc., Princeton, NJ, USA}
  \icmlaffiliation{gdm}{Google DeepMind, Mountain View, CA, USA}
  \icmlaffiliation{googlede}{Google Inc., M\"{u}nchen, Germany}

  \icmlcorrespondingauthor{Alex Mathai}{alexmathai@cs.columbia.edu}

  \icmlkeywords{LLM Agents, Fuzz Repair, Linux Kernel Repair, ICML}

  \vskip 0.3in
]




\printAffiliationsAndNotice{}


\begin{abstract}
\vspace{-1mm}
Fuzzing frameworks like syzkaller have uncovered thousands of Linux kernel crashes, many of which are critical and security-sensitive. However, the ability to rapidly repair these crashes has not kept pace, particularly given the complexity and low-level nature of kernel code.
Predominantly targeting user-space applications, existing LLM-based program repair techniques are not tailored to the unique challenges posed by kernel fuzz bugs—such as the absence of natural language bug reports, lack of exhaustive test oracles, and highly specialized crash artifacts.
Thus, in this work, we first identify the prevalent bottlenecks that generic agents struggle with in complex systems such as the Linux kernel.
Guided by these findings, we then build \agent, a workflow-based agent inspired by how kernel developers diagnose and fix bugs, and \sysplus, the co-designed toolstack supporting \agent's requests.
\agent inspects relevant execution logs, generates execution-grounded natural language hypotheses, synthesizes candidate patches, validates patches through crash reproduction, and \textit{iteratively} refines its reasoning.
We ablate these agentic system features in \agent and quantitatively analyze their contributions to the overall performance.
We also report our experience on building agents for kernel crash repair.
Although derived from this work on Linux, we note that our experience applies broadly to fuzzing-discovered bugs in complex systems software.
We evaluate \agent on \data
and show that it can repair up to 54.5\% of crashes without localization and 65\% with correct file hints. We also show \agent's generalization on a few wild Syzkaller bugs, and how different patch types offer varying utility to developers who debug complex system software.


\vspace{-5mm}

\end{abstract}

\section{Introduction}

\textbf{Target Problem.} Repairing system crashes discovered by fuzzers is a critical but largely underexplored challenge in software engineering. Kernel fuzzers such as \syzkaller~\cite{syzkaller} have proven highly effective at uncovering deep and subtle bugs in complex, low-level systems like the Linux kernel~\cite{linux}. \syzkaller uses coverage-guided fuzzing to systematically generate and mutate system-call inputs, enabling it to explore diverse kernel execution paths.
To date, it has reported over $6K$ kernel crashes (which were eventually fixed), and continues to discover more on a regular basis;
many of those crashes expose high 
risk~\cite{syzscope}, and are assigned CVE numbers~\cite{bursey2024syzretrospectorlargescaleretrospectivestudy}. 
However, while fuzzing can identify crashing inputs quickly and at scale, its practical value is limited if we cannot patch the discovered bugs with similar speed and scalability.

Recent advances in LLM-driven coding agents have shown significant promise in program repair~\cite{yang2024sweagent, zhang2024autocoderover, liu2024marscodeagent, xia2024agentlessdemystifyingllmbasedsoftware}, but most target application-level code.
Application software typically operates in user space with clearer specifications and fewer side effects, making it more amenable to automated repair.
In contrast, Linux kernel code is low-level, performance-critical, and tightly coupled to hardware. It contains implicit, often undocumented system-level invariants that surface as subtle deadlocks, data races, and privilege violations that can compromise system stability. 
In this work, we take a step toward bridging this gap by targeting the repair of Linux kernel crashes triggered by fuzzed inputs. Specifically, we not only introduce a kernel repair agent, but also discuss (i) the underlying design choices for LLM-agents in complex systems software, and (ii) the practical utility of agentic patches in such under-specified environments.  
Given a crash report, reproducing input, and the kernel codebase, our goal is to synthesize a patch that eliminates the crash while preserving system functionality. 

\vspace{-3mm}
\paragraph{Challenges.} 
Unlike application-level bugs in  popular bugfix datasets (SWE-Bench~\citep{jimenez2024swebench}), 
which often come with natural language (NL) descriptions and test cases, kernel crashes discovered by fuzzers pose several unique challenges for LLM-based repair:

\begin{enumerate}[leftmargin=*, topsep=0pt, noitemsep]

\item \textit{Lack of failure semantics} \textbf{(C1)}: Fuzzer-reported crashes lack natural-language explanations, test oracles, and failure specifications; instead the inputs are raw machine artifacts (stack traces, register dumps, sanitizer logs) that resist direct LLM reasoning.

\item \textit{Scale and execution opacity} \textbf{(C2)}: The Linux kernel spans 20M+ LOC across 50K+ files, and its runtime behavior — interrupt handlers, concurrent threads, and hardware-coupled control flow - is hard to recover from static source inspection alone.
Experienced developers bridge this gap through subsystem-specific intuition accumulated over decades, a tacit expertise that doesn't transfer via prompting to a general-purpose LLM. Thus, effective kernel repair requires both narrowing the code search space and reconstructing the dynamic execution path in a form that the agent can reason over.

\item \textit{Validation cost} \textbf{(C3)}: Verifying a patch requires a full kernel rebuild and reproducer execution — $\sim$ 30-60 minutes per attempt — prohibitively expensive for the iterative exploration that LLM-based repair demands. 
\end{enumerate}

\noindent This exposes a mismatch between what kernel repair provides and what agents need. We note that these specific challenges extend beyond fuzz-based kernel crash repair to other fuzz-based bugs (\textbf{C1}) and systems-software codebases (\textbf{C2},~\textbf{C3}). 
Closing this gap, in our experience, required more than a stronger LLM: we had to reshape the artifacts, evidence, and feedback loop into forms an agent could reason over. \textbf{Co-designing} agents (\agent) and tool-stacks (\sysplus) was what that reshaping looked like in practice.


\textbf{What we finally built.} After several iterations, we converged on a co-design of \agent and \sysplus that addresses C1–C3 as follows. Our early attempts treated this as a prompting problem — feeding raw crash artifacts to LLMs and expecting them to reason their way to a fix. They didn't. The artifacts were too machine-flavored (C1), the search space too large (C2), and the feedback loop too slow (C3) for the agent to recover. What worked, in the end, was reshaping the inputs, the search, and the loop in tandem:

\begin{enumerate}[leftmargin=*, topsep=0pt, noitemsep]
\item \textit{Hypothesis-driven reasoning} (for C1): We learned to convert raw crash artifacts into natural-language hypotheses before attempting any code edit. This gives the agent an interpretable belief state to reason and iterate over.
\item \textit{Trace minimization and targeted localization} (for C2): Rather than trying to imbue the agent with subsystem-specific expertise (which failed), \sysplus supplies structured runtime evidence — minimized execution traces aligned with the crashing stack trace and a narrowed set of localization candidates — that grounds the agent's reasoning in observed behavior rather than prior knowledge. Prior execution-guided LLM research targets small user-space programs with interpreter traces \citep{chen2018execution, zhong2024debuglikehumanlarge}. Making this tractable at Linux kernel scale, where traces span millions of events across interrupt handlers and concurrent threads, required new tooling.
\item \textit{Fast feedback} (for C3): We could not have iterated on the above without first cutting per-iteration cost. \sysplus's incremental kernel builds, and per-file syntax-filters reduce each iteration cost from $\sim20$ minutes to seconds, making \agent's test-time exploration tractable.
\end{enumerate}

\vspace{-3mm}
\paragraph{Results.} We use \sysplus to conduct over $500K$ kernel patching experiments. Our evaluation spans both \data—a benchmark of \syzkaller bugs~\cite{mathai2024kgym} and some wild \syzkaller bugs 
. On \data, \agent resolves $\sim 65\%$ of kernel crashes when given file-level localization information. Supplying a minimal, relevant execution trace further boosts performance by $\sim 9\%$. Without localization information, \agent solves $\sim 54.5\%$ of crashes. 
A manual study reveals that $32$\% of produced patches are semantically equivalent to the developers' fixes.  


\textbf{Contributions.} This paper details our experiences building agents for kernel crash repair. Our contributions are:

\begin{enumerate}[leftmargin=*, noitemsep, topsep=0pt]
 \item  \textbf{Lessons on fuzz-bug repair} (\S\ref{sec:motivation}, \S\ref{subsec:hyp}, \S\ref{subsec:patch}): We document why agents designed for human-reported bugs do not fit squarely to fuzzer-discovered crashes, and what artifact reshaping — hypothesis scaffolding (\S\ref{subsec:hyp}) and trace minimization (\S\ref{sec:execution}) — is needed to close the gap.
 \item  \textbf{Lessons on agent–tool co-design} (\S\ref{sec:execution}, \S\ref{subsec:tool_results}): We report concrete bottlenecks (context pressure, iteration latency, noisy logs) that emerge at Linux kernel scale, and the tool-side decisions (\sysplus) we made in response.
 \item  \textbf{Lessons on patch utility} (\S\ref{subsec:utility_results}, \S\ref{sec:cases}): Through manual analysis of $79$ patches, we surface a plausible/helpful/incorrect taxonomy that is more nuanced than pass-rate metrics. We also raise an open question about functionality preservation in codebases with low test coverage.
\end{enumerate}

\agent (\S\ref{subsec:hyp}–\S\ref{subsec:patch}) and \sysplus (\S\ref{sec:execution}) are the artifacts through which we derived these lessons; we describe them in detail so others can build on our experience. \S\ref{sec:evaluation} reports the empirical study that grounds the lessons above.

\begin{figure*}[t]
    \centering
    \includegraphics[width=\linewidth]{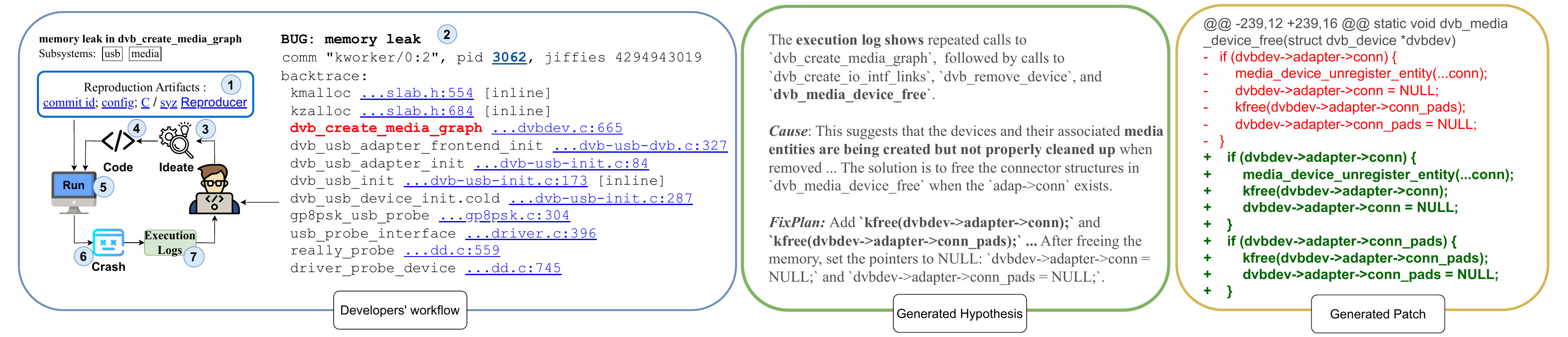}
    \caption{The crash-fixing workflow of a kernel developer. \agent follows this flow to generate {\em hypothesis} and {\em patch}.}
    \vspace{-7pt}
    \label{fig:syzkaller-bug}
\end{figure*}

\vspace{-3mm}
\section{Motivation and Overview}
\label{sec:motivation}

This section begins by describing how kernel developers typically fix crashes (\S\ref{sec:crash_fix}), which motivates the design of our system. It then presents an overview of the \agent pipeline (\S\ref{sec:kagent-overview}) and highlights key improvements in \sysplus that enable and support \agent.

\vspace{-3mm}
\subsection{Developer Crash-Resolution Workflow}
\label{sec:crash_fix}



Figure~\ref{fig:syzkaller-bug} shows how a kernel developer typically resolves a new crash, using a real \syzkaller-reported bug~\cite{syzkaller}. Upon detection, an automated report (\circled{1}) is sent to the developer mailing list, including the bug type (e.g., “memory leak”), crashing function (e.g., \texttt{dvb\_create\_media\_graph}), commit ID, build config, and reproduction scripts. The developer then inspects machine-generated diagnostics like the crash report, bug type, and stack trace; as the bugs lack natural language explanations due to their fuzzing origin (\circled{2}). Using this and their code understanding, the developer identifies relevant files, hypothesizes the root cause (\circled{3}), and modifies the code (\circled{4}). Next, the developer rebuilds the kernel and runs the reproducer script (\circled{5}) to verify the fix. Often, the initial patch does not fully resolve the issue (\circled{6}), prompting inspection of kernel logs (\circled{7}) to refine understanding of the faulty code paths. The developer then iterates—hypothesizing, patching, and testing—until the crash is resolved and the fix passes kernel maintainer review. 

This workflow highlights three capabilities that directly inform our design: \textit{iterative hypothesis testing} (developers refine hypotheses through trial and error — and LLM agents, lacking deep contextual expertise, typically require more iterations than humans), \textit{trace-guided reasoning} over minimal execution slices, and \textit{fast rebuild–test cycles}. \S\ref{subsec:hyp}-\S\ref{sec:execution} detail how we realize these capabilities at scale. 

\subsection{Method Overview}
\label{sec:kagent-overview}

\begin{figure}[t]
    \centering
    \includegraphics[width=1\linewidth]{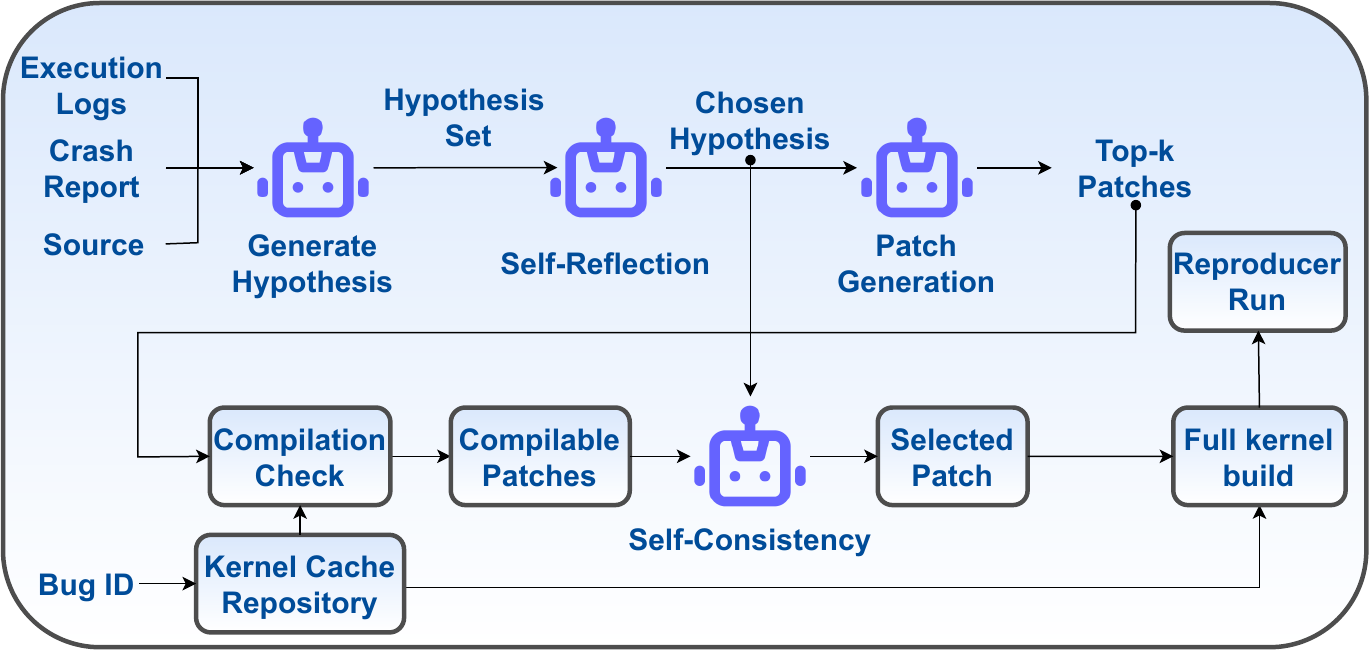}
    \caption{Method Overview.}
    \label{fig:patch_gen_select}
    \vspace{-5mm}
\end{figure}

Following these insights, Figure~\ref{fig:patch_gen_select} presents our crash resolution pipeline, with our core contributions (\agent and \sysplus). Table~\ref{tab:defnitions} summarizes key system terminology. 

\vspace{-3mm}
\paragraph{\agent. (\S\ref{subsec:hyp}-\S\ref{subsec:iter})} 
The pipeline begins with \agent receiving three key inputs: execution logs, a crash report, and relevant source code. Using trace-guided reasoning, \agent analyzes control and data flow to (i) identify plausible root causes of the fault, and (ii) generate candidate fixes. For example, in the memory leak case (Figure~\ref{fig:syzkaller-bug}), \agent localizes the bug to \texttt{dvbdev.c}, identifies a missing \texttt{kfree}, and proposes a mitigation strategy (green box). \textit{Hypothesis scaffolding} is the structured reasoning loop through which \agent iteratively tests and refines hypotheses. For each hypothesis, it generates candidate patches, filters out those that don't compile, and performs a self-consistency check to ensure the patch aligns with the hypothesis. The most consistent patch is validated via \sysplus through fast kernel builds and reproducer execution. If the patch fails, \agent uses the outcome as feedback to revise its hypothesis or patch strategy. This loop continues until a working fix is found—often matching developer-written patches. This iterative, hypothesis-driven process is key to enabling systematic and explainable kernel crash repair. 

\vspace{-3mm}
\paragraph{\sysplus. (\S\ref{sec:execution})}
\label{sec:kgym-overview}
To support scalable hypothesis-driven debugging, \sysplus must accelerate agent reasoning loops. Thus, in \sysplus, we support selective tracing: instead of capturing all kernel logs, we extract and compress a minimal, high-signal subset of function calls, allowing effective trace-based LLM reasoning. We also address inefficiencies in patch validation by enabling fast syntax checks and incremental kernel builds, dramatically reducing iteration time. These enhancements enable high-throughput kernel repair, making \sysplus critical to crash resolution.

\begin{table*}[t]
    \centering
    \scriptsize
    \caption{Terminologies used in \agent and \sysplus frameworks.}
    \label{tab:defnitions}
    \renewcommand{\arraystretch}{0.95}
    \setlength{\tabcolsep}{4pt}
    \begin{tabular}{l|p{0.78\linewidth}}
    \toprule
\textbf{Localization Candidates} & During crash resolution, we identify a small set of likely fault-related source files, termed \textit{localization candidates}. \\
         \midrule
\textbf{Reproducer Run} & Each \syzkaller bug includes a \textit{reproducer program} that, when executed on the compiled kernel, triggers the crash. We define a \textit{reproducer run} as executing this program for up to 10 minutes or until a crash occurs. \\
\midrule
\textbf{Crash Resolution} & 
The goal of crash resolution is to generate a patch that prevents the reproducer from triggering the crash.
\\ \midrule
\textbf{Execution Logs} & During a reproducer run, the kernel executes a sequence of functions---some of which are defined in the localization candidates. We refer to this sequence as the (function-granularity) \textit{execution logs} \\
\bottomrule
\end{tabular}
\vspace{-5mm}
\end{table*}

\section{Related Work}
\label{sec:related}

\paragraph{Benchmarks \& \sys.} 
Code generation is the most studied application of Code LLMs, with benchmarks ranging from single-file completion~\citep{chen2021evaluating} to cross-file~\citep{zhang2023repocoder,ding2023crosscodeeval}, repository-level~\citep{jimenez2024swebench}, and kernel-level tasks~\citep{mathai2024kgym, huang2026outrunningllmcutoffslive}. Since \agent targets kernel crashes, we evaluate it on \data. To support this, we build on \sys~\citep{mathai2024kgym}, a platform for reproducing and validating Linux kernel crashes. 
While \sys provides crash reproduction and kernel build pipelines, it lacks efficient builds and fine-grained tracing. 
Our enhanced version, \sysplus, adds scalable, trace-aware support for kernel repair. User-space tools like \textsc{debug-gym}~\citep{debuggym} address orthogonal challenges.

\paragraph{Execution History.} 
Prior work has used execution data to enhance LLM-generated code, often requiring manual annotation or prompt tuning~\citep{chen2023improving,wu2023fine}, incorporating interpreter outputs and explanations~\citep{chen2018execution,chen2023teaching,hu2024leveraging}, refining outputs via trained models~\citep{pearce2023examining,gupta2023grace}, or using the base model directly~\citep{le2022coderl,shinn2023reflexion,zhong2024debuglikehumanlarge}. \agent adopts the latter, using execution logs to guide debugging, but uniquely extends this to system-level crash analysis—making it the first to systematically study execution traces at Linux kernel scale.


\paragraph{LLMs for SE Agents.} LLM-Agents have been used for end-to-end SE, where agents invoke tools and act on feedback to automate tasks like program repair. Prior agents integrate tools for code search, patching, and testing~\citep{yang2024sweagent,arora2024masaimodulararchitecturesoftwareengineering,zhang2024autocoderover,ruan2024specrover,liu2024marscodeagent}, often targeting user-reported issues with natural language descriptions~\citep{jimenez2024swebench}. Passerine~\citep{passerine} extends to multilingual user-space bugs with sanitizer reports. In contrast, \agent targets Linux kernel crashes—low-level, concurrent bugs with limited NL context and sanitizer-generated diagnostics. 
A concurrent and complementary effort is Code Researcher~\cite{singh2025coderesearcherdeepresearch}. Unlike \agent, it operates in a non-oracle setting and prioritizes file localization. It retrieves relevant code via tool-augmented search, often at the cost of patch plausibility. In contrast, \agent analyzes plausible, executable patches and provides a detailed evaluation of their utility.

\section{Methodology}
\label{sec:hypothesis_testing}

\agent takes as input a crash report, execution log, and kernel source code, and produces a candidate patch to repair the crash. As illustrated in~\Cref{fig:patch_gen_select}, \agent structures the repair process around a reasoning scaffold centered on hypothesis generation and refinement. 
By explicitly separating hypothesis formulation, 
\agent improves interpretability and allows the agent to reason in stages, much like a human developer. To mitigate hallucinations, \agent anchors its reasoning in observable evidence (i.e., execution) and applies self-consistency checks and ranking to remove unsupported edits. The scaffold also supports iterative refinement—tracking hypotheses and outcomes across cycles.
To this end, \agent works in the following three phases:

\subsection{Hypothesis Generation \& Selection}
\label{subsec:hyp}

\paragraph{Generate Hypotheses.} 
In Step-I of~\Cref{fig:patch_gen_select}, given a kernel crash report ($\mathcal{R}$), execution trace ($\mathcal{T}$), and source code context ($\mathcal{C}$), \agent generates a set of plausible hypotheses that explain the root cause of the crash and proposes corresponding mitigation plans. Thus, the input context is: $\mathcal{I} = (\mathcal{R}, \mathcal{T}, \mathcal{C}).$ Using $\mathcal{I}$, \agent generates candidate hypotheses: $\mathcal{H} = \{ h_1, h_2, \dots, h_k \}$;
where each \(h_i\) is a tuple containing: \texttt{Cause}$_i$ (a plausible explanation for the observed crash) and \texttt{FixPlan}$_i$ (an outlined strategy to resolve the issue, like modifying a specific function or adding a conditional guard).
These hypotheses serve as structured inputs for downstream patch generation, guiding reasoning toward targeted code edits consistent with the $Cause_i$. 

\paragraph{Select Hypothesis.} LLMs are prone to \textit{hallucination}---producing confident but incorrect hypotheses based on fabricated or irrelevant evidence. This behavior, well-documented in tasks such as question answering and information retrieval~\cite{Huang_2025}, can derail debugging \& repair by diverting attention to unrelated code regions. To mitigate this, we incorporate a \textit{self-reflection} step. After generating $\mathcal{H}$, \agent scores each hypothesis by reviewing its consistency with $\mathcal{I} = (\mathcal{R}, \mathcal{T}, \mathcal{C})$, i.e., the crash report, execution trace, and code context. The consistency measures we use are (i) \textit{Trace alignment}: Is the hypothesis supported by the observed call stack and execution behavior?; (ii) \textit{Source locality}: Does it refer to functions and variables present in the candidate code region?; and (iii) \textit{Causal plausibility}: Does the explanation logically precede the failure in $\mathcal{R}$?

\agent then selects the top-$k$ scoring hypotheses for downstream patch generation: $ \mathcal{H}_k = \mathrm{TopK}\bigl(\{h_i\}_{i=1}^k,\,k\bigr)$.
While self-reflection is not immune to LLM error, combining generation with structured critique and ranking significantly improves robustness. This mirrors how developers often approach debugging by considering competing explanations and refining their understanding through critical evaluation.



\subsection{Patch Generation \& Selection}
\label{subsec:patch}

\paragraph{Generate Patch} 
Given a selected hypothesis $h^* \in \mathcal{H}_k$, \agent generates code edits to realize $h^*$. \agent generates $\mathcal{P} = \{p_1, \dots, p_m\}$ which is a set of candidate patches ($p_j$). For each $p_j$, rather than producing a complete \texttt{git diff}---which is highly sensitive to line numbers and surrounding context---we adopt a \textit{replace}-based interface. Thus each $p_i$ is a list of edits of the form: $\mathcal{E} = \{\, e_1, e_2, \dots, e_n\,\}$,
where each $e_i$ is a tuple containing a snippet of existing source code, the proposed change, and a natural language justification. Inspired by prior works~\cite{zhang2024autocoderoverautonomousprogramimprovement, xia2024agentlessdemystifyingllmbasedsoftware}, this interface is robust to patch-formatting issues and clarifies each edits' intent.
\agent then submits each patch $p_j$ to \sysplus for fast compilation filtering. Thus, each patch $p_j$ is tested for compilability 
using the cached kernel configuration (\cache) associated with the crash. Uncompilable patches are discarded, and the final set of compilable patches ($\mathcal{P}_\text{valid}$) is saved.

\vspace{-3mm}
\paragraph{Patch Selection.}  
From the surviving patches $\mathcal{P}_\text{valid}$, \agent selects $p^*$, the patch most aligned with the original hypothesis $h^*$. This step, called \emph{self-consistency reasoning}, prompts \agent to evaluate each patch $p_j$ with respect to its coherence with the cause and fix plan of $h^*$.
This mimics internal code review, encouraging \agent to prefer patches that faithfully realize the hypothesis.


\vspace{-3mm}
\paragraph{Termination Criteria.}  
The hypothesis scaffolding loop terminates when either a valid patch successfully resolves the crash or all top-$k$ hypotheses have been exhausted. If no patch candidate under a given hypothesis $h^*$ passes the compilation filter or aligns with the intended fix, \agent deems $h^*$ invalid and proceeds to the next hypothesis in $\mathcal{H}_k$. If a consistent, compilable patch $p^*$ is found, it is forwarded to \sysplus for a full kernel build and reproducer execution. 

\vspace{-3mm}
\paragraph{Iterative Debug \& Repair}
\label{subsec:iter}

Like human developers, \agent iterates to identify root causes and effective patches, using failed attempts as feedback. To avoid redundant steps, it tracks past hypotheses, code states, and outcomes. To scale this process, \agent balances exploration and exploitation. Exploitation occurs when \agent performs long-range trajectories involving multiple sequential patch-build-test (or \textit{debug cycle}) attempts. Thus, exploitation is measured in $\mathcal{D}$-the number of debug cycles in the trajectory. On the other hand, exploration occurs when \agent spawns multiple branches from a shared state. Each branch is a distinct hypothesis, creating a tree-like search, wherein each trajectory is a root-to-leaf path. Thus, exploration is measured in $\mathcal{B}$-the branching factor of the tree.

\begin{figure}[t]
    \centering
    \includegraphics[width=\columnwidth]{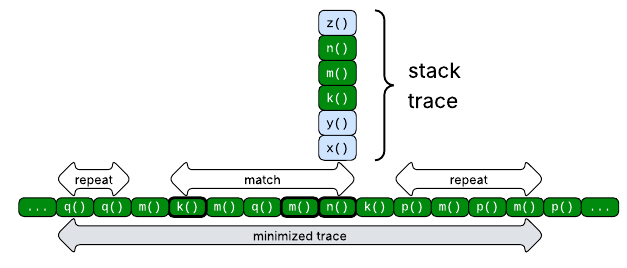}
    \caption{
    The execution trace-minimization algorithm starts by scanning the execution of all traced functions (green) backwards to find all stack-trace functions from the localized file ``in order''.  In this example, the stack-trace has functions n, m, and k-defined in the localized file. Other untraced functions (blue) are ignored. This matching region, labeled “match”, may also include additional functions like \texttt{q} (defined in the localized file, present in the execution trace, but absent in the stack trace) and repeated calls like \texttt{m}. We refer to this first step as ``stack trace anchoring''. Then, the span is expanded backwards and forwards until a repeating pattern of traced functions is detected (\textit{\{q\}} in the prefix, \textit{\{p, m\}} in the suffix), or up to a maximum number of function records.
    }
    \label{fig:execution_minimization}
\vspace{-5mm}
\end{figure}

\subsection{Tool Support for Scalable Crash Repair}
\label{sec:execution}

To scale LLM-driven debugging \& repair, we embrace the paradigm of agent–tool co-design. 
\textit{Tools affect Agents:} Rather than relying on general-purpose tools, we design \agent to invoke
dedicated kernel-specific tooling that precisely exposes the information needed for crash repair.
\textit{Agents affect Tools:} Conversely, tool design is guided by agent capabilities and bottlenecks, such as context length, reasoning reliability, and iteration speed. This co-design loop allows us to unlock capabilities that neither agents nor tools achieve in isolation. To this end, we introduce \sysplus, a tool framework that enables two critical optimizations for kernel debugging: 
(1) condensing kernel execution traces into focused, high-signal subsets for trace-based LLM reasoning, and 
(2) accelerating patch validation via incremental builds and fast compile-time checks.

\vspace{-3mm}
\paragraph{Identifying Minimal Execution Traces}
Full kernel logs are often noisy and exceed LLM context limits (at times $\sim 100K$ entries), while stack traces alone may miss root causes. Hence, there is a need to extract \textit{minimal} and \textit{high-quality} signals from these traces. As part of its agent tooling, \sysplus instruments candidate source functions and collects execution logs using a lightweight in-memory tracer based on \tprintk~\cite{Rostedt}. Upon a crash, \sysplus retrieves the kernel's execution traces via \texttt{kDump}. 

To improve the signal-to-noise ratio for agent reasoning, we introduce a trace minimization algorithm built into \sysplus. It follows two principles: (1) isolate the crashing thread using the process identifier (or PID), and (2) extract a concise trace window centered around the crash stack and the suspected source file (if known). The algorithm filters traces to retain entries only from the crashing thread. It then applies \textit{stack trace anchoring} (see Figure~\ref{fig:execution_minimization}), scanning backward to align observed frames. From the anchor, it expands in both directions up to a fixed budget (200 functions) or until repeating call patterns emerge. For the memory-leak example in Figure \ref{fig:syzkaller-bug}, the trace (Figure \ref{fig:minimized_execution}) anchors at \texttt{dvb\_create\_media\_graph}, and includes \texttt{dvb\_media\_device\_free}, which is \textit{inside} the trace (but \textit{outside} the stack) and is the correct fix location.

\vspace{-3mm}
\paragraph{Efficient Kernel Builds}
\label{sec:iterative_edits}
Fast iteration is crucial for agent-driven debugging. Thus, \sysplus incorporates a kernel-build tool that leverages \cache, an object cache storing compilation artifacts across sessions. When a patch is proposed, \textit{only modified files} are recompiled, and \cache is reused for the rest, reducing build times from $19$ minutes to under three minutes on 8-vCPU systems. Additionally, for uncompilable patches, \sysplus adds a rapid syntax-checking step using per-file \texttt{make} commands to prevent unnecessary ``complete build'' attempts. This filters invalid candidates within 2-3 seconds, ensuring only valid patches proceed to full validation. These optimizations make agent-guided debugging tractable even on large kernel codebases.

\vspace{-3mm}
\section{Results}
\label{sec:evaluation}

\textbf{Hardware.} \sysplus is deployed on Google Cloud Platform using two clusters, each with $37$ compute engines having $88$-vCPUs. To date, \sysplus has conducted over $500K$ kernel-build and kernel-patching experiments.

\paragraph{Datasets.} We evaluated \agent on \data~\citep{mathai2024kgym}, a curated benchmark of kernel bugs. 
Each bug has metadata, reproducers, developer fixes, and crash reports. Following \citet{mathai2024kgym}, we partition \data into: (a) a small-context subset (SCS, $117$ bugs) where localized files fit within $16K$ tokens, and (b) a large-context subset (LCS, 162 bugs) requiring up to $50K$ tokens. This partitioning enables fast iteration on SCS and scalable evaluation on LCS. We also report results on a few wild \syzkaller bugs (\Cref{sec:cases}).

\paragraph{LLMs \& Baselines.} Our methodology is LLM-agnostic. For cost efficiency, we use Gemini 1.5 Pro-001 for preliminary studies and ablations. We report our best results 
using Gemini 2.5 Pro and 1.5-Pro-002. Our baseline \cite{mathai2024kgym} predicts $10$ patches to try and resolve a kernel bug.


\paragraph{Research Questions.} We structure our experiment section into the following research questions:

\begin{itemize}[leftmargin=*,topsep=0pt]
    \item \textbf{RQ1}: How effective is \agent at resolving Linux kernel crashes? How much does each component contribute?
    \item \textbf{RQ2}: How useful are \agent patches? 
    \item \textbf{RQ3}: How well can \agent localize bugs?
    \item \textbf{RQ4}: Is \sysplus effective for scalable crash repair?
    \item \textbf{RQ5}: What is the cost and latency of \agent?
\end{itemize}

\begin{table*}[!htbp]
\centering
\scriptsize
\renewcommand{\arraystretch}{1.0}
\setlength{\tabcolsep}{4pt}
\caption{~Average crashes resolved in $3$ runs (SR: Self-reflection; SC: Self-consistency; \textit{Traj}: number of trajectories; default results are with Gemini-1.5-pro-001; ** indicates 1.5-pro-002; \#\# indicates 2.5-pro~; $\mathcal{D}$ and $\mathcal{B}$ are depth and branching factor of each trajectory
}
\label{tab:kbench_results}
\begin{tabular}{p{0.6\linewidth}|rrr|r}
\toprule
\textbf{Strategy} & \textit{Traj} & \textbf{$\mathcal{D}$} & \multicolumn{1}{c|}{$\mathcal{B}$} & \begin{tabular}[c]{@{}c@{}}\textbf{Average Crashes Resolved} \textbf{(min--max)}\end{tabular} \\ \midrule
\multicolumn{5}{c}{\textbf{Small Context Subset of 117 bugs (SCS) : The bugs with a context length $<$ 16K tokens}} \\ \midrule
\#1. ~\textbf{Baseline Top-10} :  kGym \cite{mathai2024kgym} Baseline (Top-10 predictions)  & 10 & 1 & {1} & 10 \\ \midrule
\#2. ~\textbf{Edit-Compile-Test} : Simple patch-build-test loop with edit-level reasoning to justify each edit.   & 1 & 4 & 2 & 42 (42--42) \\ \midrule
\#3.~\textbf{Hyp then Patch} : Single hypothesis $\rightarrow$ generate multiple patches $\rightarrow$  select first compilable patch.   & 1 & 4 & {2} & 46.33 (39--51) \\ \midrule
\#4.~\textbf{Hyp-SR then Patch} : Extending Strategy~\#3 to generate {\em multiple} hypotheses (Hyp) and select the best with self reflection (SR)   & 1 & 4 & {2} & 48.33 (47--49) \\ \midrule
\#5.~\textbf{Hyp-SR then Patch $\mathcal{F}$} : Extending Strategy~\#4 with several narrow trajectories ($\mathcal{B}=1$), rather than a tree-search with a high branching factor ($\mathcal{B}=2$)   & 4 & 4 & {1} & 54.67 (51--57) \\ \midrule
\#6.~\textbf{~Hyp-SR then Patch-SC $\mathcal{F}$} : Extending Strategy~\#5 with patch self-consistency (SC); the agent chooses the patch most consistent with the generated hypothesis.   & 4 & 4 & 1 & \textbf{58.33} (57--59) \\ \midrule
\quad\quad\quad\quad- 7~Strategy~\#6 with Gemini-1.5-pro-002  ** & 4 & 4 & {1} & \textbf{62.33} (60--64) \\ \midrule
\quad\quad\quad\quad- 8~Strategy~\#6 with with Gemini-2.5-pro \#\# & 4 & 4 & {1} & \textbf{76.67} (74--79) \\ \midrule
\multicolumn{5}{c}{\textbf{Large Context Subset of 162 bugs (LCS) : The bugs with a context length up to 50K tokens}} \\ \midrule
\quad\quad\quad\quad- 9~Strategy~\#6 on LCS with Gemini-1.5-pro-001   & 4 & 4 & {1} & \textbf{65.33} (61--71) \\ \midrule
\quad\quad\quad\quad- 10~Strategy~\#6 on LCS with Gemini-1.5-pro-002  ** & 4 & 4 & {1} & \textbf{75} (72--79) \\ \midrule
\quad\quad\quad\quad- 11~Strategy~\#6 on LCS with Gemini-2.5-pro \#\# & 4 & 4 & {1} & \textbf{101} (100--102) \\ \bottomrule
\end{tabular}
\vspace{-3mm}
\end{table*}


\subsection{RQ1.~\agent's Crash Repair Effectiveness}

\textbf{Setup.} In this RQ, we run experiments in the oracle setting, i.e., we provide the oracle buggy file to \agent; 
This setting enables a more focused assessment of \agent’s crash-repair capabilities, as it reflects a realistic human-in-the-loop workflow—as developers often have a tentative hypothesis about the faulty file and use automated tools to generate a candidate fix. It also aligns with common assumptions in prior bug-repair literature~\cite{sargam-tse,xia2022less, jiang2023impact}. Patch success is measured by the number of crashes \agent can resolve.

\textbf{Results.} \Cref{tab:kbench_results} details \agent's results on the \data dataset. Row $7$ shows that \agent with Gemini 1.5-pro-001 resolves $\sim 58$ SCS crashes—5.8$\times$ more than the \sys baseline. It also resolves $\sim 65$ bugs of LCS crashes (row $10$). With progressively stronger models (Gemini 1.5-pro-002 and Gemini 2.5 pro), mitigated crashes increase steadily to $\sim 76$ (SCS) and $\sim 101$ (LCS) (rows $8$ \& $11$). Table \ref{tab:kbench_results} ablates different settings with \agent (see below). 

\begin{itemize}[leftmargin=*, topsep=0pt, noitemsep]
    \item \textit{Edit-Compile-Test} (\Cref{tab:kbench_results}: strategy 2). The basic Edit-Compile-Test loop resolves 42 Linux crashes by using (original, replaced) pairs with edit-level explanations.


    \item \textit{Hypothesis Generation then Patch Generation} (\Cref{tab:kbench_results}: strategy 3 \& 4). In this strategy, \agent: (i) generates a hypothesis, (ii) creates candidate patches, (iii) filters out the uncompilable ones, and (iv) builds and tests the first valid patch. This resolves $\sim 46$ bugs, increasing to $\sim 48$ with self-reflection. \agent uses temperature 0.8 for diverse hypotheses/patches, and 0.2 for other steps.

   \item \textit{ $\downarrow \mathcal{B}$ and $\uparrow$ Traj} (\Cref{tab:kbench_results}: strategy 5-11). We see clear gains by decomposing the original \textit{broad} tree ($\mathcal{B}=2$) into many narrow trajectories. Applying the \textit{Hyp then Patch} strategy with 4 trajectories (\(\mathcal{D}=4, \mathcal{B}=1\)) raises performance to \(\sim 54\) resolved bugs. This improvement stems from giving \agent multiple fresh starts, avoiding early edits that adversely affect downstream actions. Adding self-consistency to select the patch best aligned with the generated hypothesis further boosts performance to \(\sim 58\). This trend continues for Gemini-1.5-pro-002 \(\sim 62\) and finally the best model Gemini-2.5-pro  \(\sim 76\).
    
\end{itemize}

\vspace{-2mm}
\textbf{Adding Execution Traces.} We extract execution traces for $255$ out of $279$ \data 
bugs with the remaining $24$ failing due to kDump misconfigurations or unsupported Debian VM images. Thus, we evaluate the effectiveness of adding execution traces on these $255$ bugs (Table \ref{tab:kbench_with_execution_results}). In LCS, \agent has relative improvements of $14.6\%$ using 2.5-pro, $10.36\%$ using 1.5-pro-002, and $13.61\%$ using 1.5-pro-001. In SCS, we see a relative increase of $9.7\%$ with 1.5-pro-002 and consistent results for 1.5-pro-001 and 2.5-pro.

\textbf{Multi-turn.} To understand the importance of multiple edit attempts, we measure the \% of bugs only solved at depths greater than $1$ in SCS. For 1.5-pro-001, this accounts for $\sim 50\%$ of solved bugs in our final strategy (without execution) and $\sim 37.56\%$ with execution, demonstrating that tree-based exploration is broadly effective within the benchmark. The corresponding numbers for 1.5-pro-002 are $35\%$ \& $27\%$ and for 2.5-pro are $33\%$ \& $24\%$ respectively.


\vspace{-2mm}
\subsection{RQ2.~Utility of \agent patches}
\label{subsec:utility_results}
\textbf{Setup.} Many plausible solutions can resolve a kernel crash, so predicting a patch equivalent to a developer's is challenging. However, a generated patch may resolve the crash while compromising functionality. Thus, we manually analyze the patches for $79$ bugs in one run of Gemini 2.5-pro on the SCS (with execution) to measure the utility of the \agent patches compared to the ground-truth patches.

\textbf{Results.} We find that patches fall into three categories. $(1)$ We consider patches as \textit{plausible} if they are semantically equivalent to the developer patch. $26$ out of $79$ \agent patches are plausible (32.91\%). $(2)$ We consider a patch as \textit{helpful} if it targets the root cause correctly, but is not perfectly equivalent. $12$ \agent patches are helpful (15.18\%). $(3)$ An \textit{incorrect} patch resolves the crash but in the process modifies (or \emph{amputates}) functionality. 
$41$ \agent patches are incorrect (51.89\%), raising an interesting open research question-``how can agents verify that \textbf{\textit{intended functionality is preserved}}, after a crash is mitigated?''--especially relevant for codebases with low test coverage (like Linux). We show examples of these categories in \Cref{sec:cases}.





\vspace{-3mm}
\subsection{RQ3. Bug Localization}

\textbf{Setup.} We consider the problem of localization at the file granularity. Localizing buggy files in the Linux kernel is challenging due to its massive scale (50K+ files). Since we are focusing on crash bugs, we adopt a simple localization strategy - we assume the buggy files appear in the crash report. This heuristic is particularly true for Linux (as most fixes are single-file) and for \data (74\% of samples modify a single file mentioned in the report). 

For this subtask, \agent is given the crash report and identifies the top-$k$ most likely buggy files. We then compare its predictions against the ground-truth files modified by the developer patch, computing accuracy as the intersection between the predicted buggy file and actual modified files. For single-file fixes, this is straightforward, as the presence of the buggy file in \agent's prediction indicates a perfect score. However, for multi-file fixes, it is possible for \agent to predict a subset of the buggy files (not \emph{all} the developer-modified files). As such, in these cases, we compute the recall of \agent's predictions, i.e., the set of correctly predicted files out of all developer modified files.

\begin{table}[t]
\centering
\caption{Buggy file prediction accuracy from the crash report. *Crash report includes correct file for 207 out of 236 single-file bugs and at least one correct candidate for 37  out of 43 multi-file bugs. Thus, we predict accuracy among these files.} 
\label{tab:oracle_file_pred}
\resizebox{\linewidth}{!}{
\begin{tabular}{c|c|cccc|cc}
\toprule
   & & \multicolumn{4}{c|}{Top-K} &    \\ 
\multicolumn{1}{c|}{Metric} & \multicolumn{1}{c|}{Bug Type} & 1 &  3 &  5 & 7 & \begin{tabular}[c]{@{}c@{}}Total\\ \# of Bugs*\end{tabular}  \\ \toprule

{Intersection (max=207)} & Single-File  & 104  & 135   & 141  & 142 & 207  \\ \midrule

{Recall (max=1)}       & Multi-File   & 0.4   & 0.48  & 0.53  & 0.55 & 37  \\ \bottomrule
\end{tabular}
}
\vspace{-5mm}
\end{table}

\textbf{Results.} Table~\ref{tab:oracle_file_pred} shows \agent's ability to predict the correct localization candidate from a crash report. For 207 of 236 single-file bugs, the developer-modified file appears in the report. As shown, \agent’s prediction improves from 104 to 142 as the number of candidate files increases. Our best model (Gemini-2.5-pro) which fixes 158 bugs out of all 236 single-file bugs, yields a total of 143 successful fixes on this 207-bug subset . Thus, when the relevant file is present, patch success is comparable in oracle ($143$) and non-oracle ($142$) settings. Misses occur when the file is absent from the crash report.

For multi-file bugs, $37$ (from $43$ in total) have mentions of localization candidates in the crash report. In Table \ref{tab:oracle_file_pred}, we depict the average recall of the predicted set of localization candidates (from the actual set of ground truth files). As shown, \agent achieves recall scores from $0.4$ to $0.55$. Our best-performing model (Gemini-2.5-pro), resolves $27$ bugs from these $37$ multi-file bugs in the oracle setting. Using our file predictions for the multi-file bugs, we can fix $4$ ($K=1$) to $10$ ($K=7$) bugs in the non-oracle setting.


Hence, in the non-oracle setting with the bug-localization agent, we patch 142 out of 236 (\(60.16\%\)) single-file crashes and 10 out of 43 (\(23.25\%\)) multi-file crashes, achieving an average success rate of \(54.48\%\)—slightly below the \(64.8\%\) success rate in the oracle setting.

\vspace{-3mm}
\subsection{RQ4. Efficacy of our tool support tailored for scalable crash repair}
\label{subsec:tool_results}
\textbf{Setup.}  We empirically assess each step of the trace minimization algorithm using execution data from all \data bugs. We use the \emph{Complete Intersection Score} (CIS), which counts how often \textit{all} developer-modified functions appear in the minimized trace. Thus, CIS is binary, assigning $1$ when all relevant functions are present and $0$ otherwise.  

\textbf{Trace Minimization Tool.} Out of the 255 bugs for which we could extract execution traces, stack-trace-only functions yields a CIS of 104/255; stack anchoring improves this to 108/255, and backward/forward expansion raises it to 117 and 127 out of 255, respectively. \textit{Note}: Collecting the full PID trace without minimization achieves a CIS of 137 but exceeds LLM limits (2400 entries vs.\ our average of 12). This analysis assumes ideal localization and focuses on single-file fixes, demonstrating the effectiveness of minimization within our pipeline. 

\textbf{Optimized Build tool.} When using \sysplus on a 88-vCPU machine, the average compilation time for a kernel reduces by over $40\%$ when compared to the vanilla \sys. For more details, please refer to \S\ref{sec:kernel_build_metrics}.

\vspace{-3mm}
\subsection{RQ5. \agent - Cost \& Latency}
\textbf{Cost.} For a single \data bug, with $4$ trees (having $4$ nodes each), we complete a maximum of $16$ debug cycles. Assuming the localized file content is $100K$ tokens, and that \agent generates $8K$ tokens in each step of the pipeline, we estimate the average cost to be $\sim\$13$ for each bug. This estimate assumes that the LLM is always successful at getting a compilable patch without having to repeat the hypothesis-to-patch decomposition. In practice, this isn't true and will vary for each LLM. For 2.5-pro, we spend $\sim \$ 21.62$ per bug (on average) in the SCS, after accounting for such retries. 

\textbf{Latency.} Each debug cycle can take a maximum of $\sim15$ minutes. However, as each tree can run in parallel, the average bug should ideally take $\sim1$ hour to complete in the worst case, given enough machines.

\vspace{-3mm}

\section{Case Studies on \agent Patches} 
\label{sec:cases}

This section shows example patches generated by \agent, categorized into three types: (1) plausible patches, (2) helpful but inaccurate patches, and (3) patches that resolve the crash at the cost of modifying (or amputating) functionality. They are for both open and \data bugs.

\subsection{Plausible Patches}

We collect wild Syzkaller bugs analogous to \data (\S\ref{sec:wild_syzkaller}). So far, we have validated two crash fixes and present one of them (see~\Cref{ubsanbugsolved}).



\noindent \textbf{(1) UBSAN: shift-out-of-bounds bug.} 
Shown in Listing~\ref{ubsanbugsolved}, this bug~\cite{ubsanbug} was reported by UBSAN \cite{ubsansanitizer} in the Wacom USB device devier. The \syzkaller reproducer calls \texttt{wacom\_\-parse\_\-and\_\-register} with \texttt{features->pktlen} (i.e. packet length) set to $0$. Function \texttt{\_\_kfifo\_alloc} later calculates the size of the required memory by calling \texttt{roundup\_pow\_of\_two} with the value of \texttt{pktlen}, which unfortunately caused a ``shift-out-of-bounds'' error.

\begin{listing}[htb]
\begin{minted}{c}
static int wacom_parse_and_register(struct wacom *wacom, bool wireless)
{  ...
   features->pktlen = wacom_compute_pktlen(hdev);
   // ==> Begin Human Fix <==
+  if (!features->pktlen)
+      return -ENODEV;
   // ==> End Human Fix <==
   ...
>  error = wacom_devm_kfifo_alloc(wacom);
   if (error)
       goto fail;
   ...
fail:
   wacom_release_resources(wacom);
   return error;
}
static int wacom_devm_kfifo_alloc(struct wacom *wacom)
{
    struct wacom_wac *wacom_wac = &wacom->wacom_wac;
    // ==> Begin kAgent Fix <==
-   int fifo_size = min(PAGE_SIZE, 10 * wacom_wac->features.pktlen);
+   int fifo_size;
    ...
+   if (wacom_wac->features.pktlen == 0) {
+       hid_err(wacom->hdev, "pktlen is zero. Cannot allocate fifo.\n");
+       return -EINVAL;
+   }
+   fifo_size = min(PAGE_SIZE, 10 * wacom_wac->features.pktlen);
    // ==> End kAgent Fix <==
   ...
   error = kfifo_alloc(pen_fifo, fifo_size, GFP_KERNEL);
}
int __kfifo_alloc(struct __kfifo *fifo, unsigned int size,
   size_t esize, gfp_t gfp_mask)
{
>   size = roundup_pow_of_two(size); // ERROR !!!
\end{minted}
    \caption{UBSAN Shift-Out-of-Bounds Bug. Lines prefixed with \texttt{-} are from the buggy version; \texttt{+} indicates fixed lines.
}
    \label{ubsanbugsolved}
    \vspace{-4mm}
\end{listing}



\begin{figure}[htb]
    \vspace{-2mm}
    \centering
    \includegraphics[width=\linewidth]{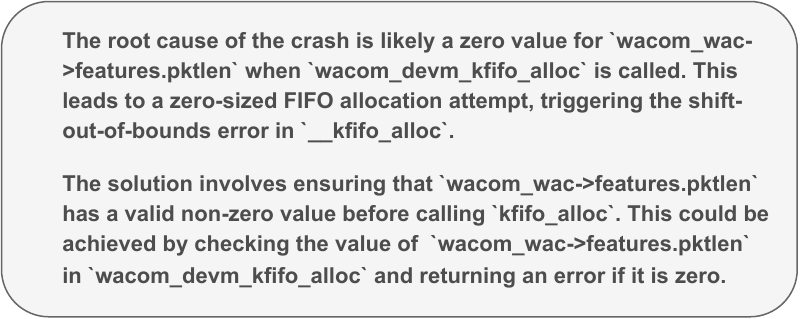}
    \caption{Hypothesis for UBSAN bug.}
    \label{fig:wacom_hyp}
    \vspace{-1mm}
\end{figure}

\textbf{\agent Fix.} 
\agent resolved this crash at level three of a debug tree. Its patch (lines $20$ to $29$) modifies function ``wacom\_devm\_kfifo\_alloc'' to return early with an error code when \texttt{pktlen} is $0$ (line $26$). This error is then handled by ``wacom\_parse\_and\_register'' and the control goes to an error handling label on line $13$ (\texttt{fail:}) that returns early from this function.
Figure \ref{fig:wacom_hyp} shows the \agent generated hypothesis.

\textbf{Developer Discussions.} Concurrently to our experiment, developers discussed this bug on the kernel mailing list.
From the discussion, we confirm that the \agent hypothesis and patch match developer understanding. Listing \ref{ubsanbugsolved} also shows the developers' suggested fix (lines $4$--$7$).

\textbf{Differences.} \agent modifies a different function from the developer, though the error-handling logic is identical. \agent returns a different error code (\texttt{EINVAL}) from the developer (\texttt{ENODEV}). Since boundaries between error codes are not strict, \texttt{EINVAL} is also applicable to the invalid argument.

\vspace{-3mm}

{\scriptsize
\begin{listing}
\begin{minted}{c}
// Developer Patch
@@ -214,7 +214,7 @@ static int erofs_fill_symlink(struct inode *inode, void *kaddr,
 	/* if it cannot be handled with fast symlink scheme */
 	if (vi->datalayout != EROFS_INODE_FLAT_INLINE ||
-	    inode->i_size >= EROFS_BLKSIZ) {
+	    inode->i_size >= EROFS_BLKSIZ || inode->i_size < 0) {
 		inode->i_op = &erofs_symlink_iops;
 		return 0;
 	}
// kAgent Patch
@@ -219,9 +219,21 @@ static int erofs_fill_symlink(struct inode *inode, void *kaddr,
        return 0;
    }
+	if (inode->i_size >= SIZE_MAX - 1) {
+		erofs_err(inode->i_sb,
+		   "symlink target size too large\n");
+		return -ENAMETOOLONG;
+	} else if (inode->i_size < 0) {
+		erofs_err(inode->i_sb,
+			"symlink target size invalid\n");
+		return -EINVAL;
+	}
    lnk = kmalloc(inode->i_size + 1, GFP_KERNEL);
-	if (!lnk)
+	if (!lnk) {
+		erofs_err(inode->i_sb,
+	      "failed to allocate memory for symlink target\n");
        return -ENOMEM;
+	}
\end{minted}
    \caption{Helpful Patch for Warning Bug\cite{warning_erofs}.}
    \label{helpful_patch}
    \vspace{-2mm}
\end{listing}
}

\subsection{Helpful Patches}

The second class of patches are those that are helpful but not perfect. These patches allow developers to understand what ``could'' work to help prevent the kernel crash but might add unnecessary or remove useful code in the process. Listing \ref{helpful_patch} shows a \data bug \cite{warning_erofs} and its developer (lines $5$--$6$) and \agent (lines $14$--$29$) patches. The developer patch adds a condition to check if \texttt{inode->i\_size < $0$} to ensure that a given symbolic link is valid. In contrast, the \agent patch does something similar but is not identical. It adds two new \texttt{if} conditions that check for both  \texttt{< $0$} and \texttt{>= SIZE\_MAX-1} on lines $18$ and $14$, respectively. However, it forgets to set the right function pointer (like line $7$) for this corner case, adds a debug statement (line $26$), and returns different error codes (lines $17$ \& $21$).    

{\scriptsize
\begin{listing}
\begin{minted}{c}
 	default:
-		WARN_ON_ONCE(1);
-		break;
-	}
-	if (!timeouts)
+		/* Unsupported protocol */
+		ret = -EOPNOTSUPP;
 		goto err;
+	}
\end{minted}
    \caption{Incorrect Patch for Warning Bug\cite{warning_cttimeout}.}
    \label{amputation_patch}
\end{listing}
}

\subsection{Incorrect Patches}
Finally, the third class of patches are those that resolve the crash but at the expense of amputating functionality. We show an example of such a patch to a \data bug that reaches a WARN call \cite{warning_cttimeout}.
In Listing \ref{amputation_patch}, \agent resolves this crash by deleting the \texttt{WARN\_ON\_ONCE} call on line $2$ and merging error handling into the default case of a \texttt{switch} statement. In contrast, the developer patch \cite{fix_commit_warning_cttimeout} adds a case to the switch statement, and modifies the code for future requirements.

\section{Conclusions}

This paper reports our experience building \agent, a crash-repair agent for the Linux kernel. We documented our agentic design 
tailored to fuzz-discovered crashes, the agent–tool co-design bottlenecks we hit at kernel scale, and a patch-utility taxonomy we used to measure utility. We hope these lessons help others build better agents for complex systems software, and we highlight open RQs, such as verifying that intended functionality is preserved after a crash is mitigated.

\bibliography{main}
\bibliographystyle{icml2026}

\newpage
\section{Appendix}
\label{sec:appendix_sec}

\appendix

\section{Collecting Wild Syzkaller Bugs}
\label{sec:wild_syzkaller}

To study plausible patches, we scraped the \syzkaller dashboard for bugs that (1) were first reported after March 10\textsuperscript{th}, 2025, (2) included valid reproducers, and (3) remained open as of April 10\textsuperscript{th}, 2025—yielding 42 bugs, with \sysplus successfully reproducing ~30. 
We conservatively treat all crash-report files (excluding frequent ones) as localization candidates, averaging $\sim$13 files per bug. This setup forms a dataset analogous to \kbenchsyz. 



While running \agent on this dataset, we encountered compatibility issues between recent \syzkaller versions and the latest Linux kernel~\cite{unregister_netdevice}, which led to unrelated kernel crashes even after the original bugs were resolved. These issues disrupted our automated pipeline. As we work to address them, we present one plausible patch generated by \agent and analyze it in detail.

\section{Details of Kernel Builds on \sysplus}
\label{sec:kernel_build_metrics}

\begin{figure}[htb]
    \centering
    \includegraphics[width=\linewidth]{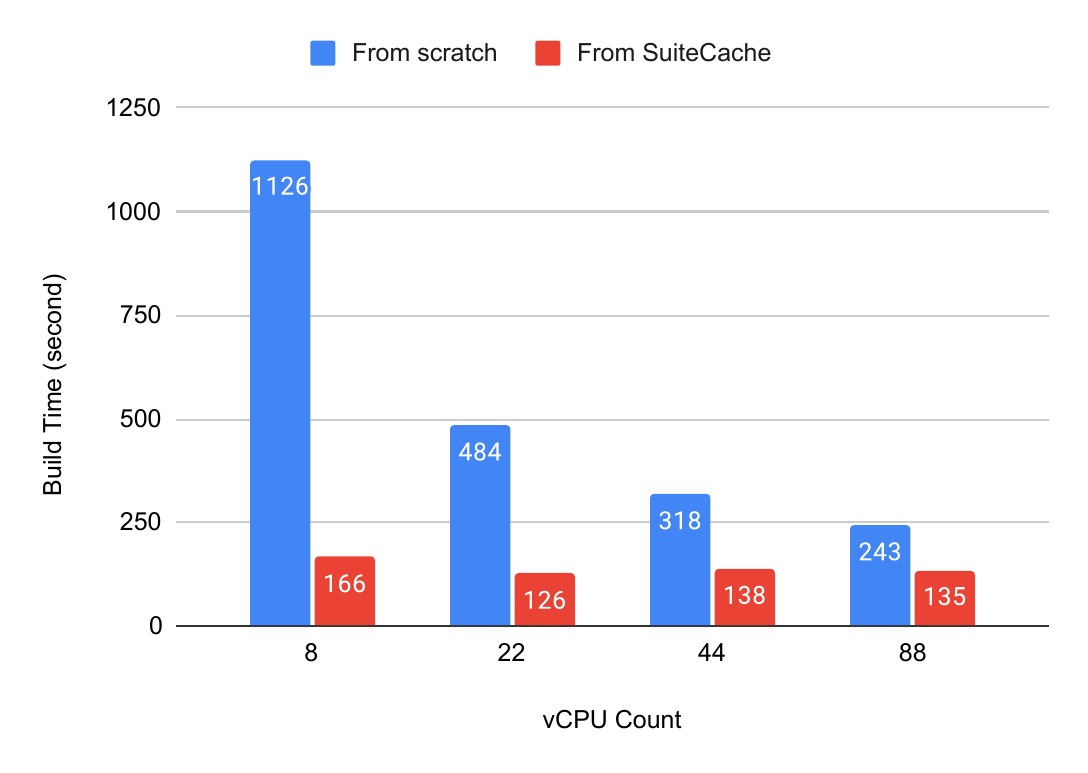}
    \caption{Kernel Build Time (from scratch vs from \cache) on a sample \data data point}
    \label{fig:build-time-vs-core}
\end{figure}

\textbf{\sys build cycles.} The \sys platform performs the following steps when a user issues a kernel-build job. 

(i) \sys checks if the user-specified Linux kernel tree (e.g. \texttt{torvalds}, \texttt{net} or \texttt{linux-next}) is available in its local Linux-repository cache. If unavailable, \sys will \texttt{git pull} the corresponding remote tree into this cache and clone a repository copy into its working directory. 

(ii) \sys will then \texttt{git checkout} the given user-specified commit and attempt to apply the user patch (specified as a \texttt{git diff} patch). If the patch application succeeds, \sys applies additional (and necessary) ``backport'' commits to resolve issues that arise from incompatibilities between old kernel source code and new compilation toolchains. 

(iii) Finally, \sys then invokes GNU \texttt{make} to build the kernel. Once the compilation is done, \sys uploads the built binaries to a cloud storage bucket.

As mentioned, the existing kernel-build process in \sys is slow as it compiles the entire kernel from scratch regardless of the number of files the patch modifies. As shown in Figure \ref{fig:build-time-vs-core}, the above-detailed process can take a significant amount of time to complete. As depicted, when building from scratch using \sys, a \kbenchsyz bug can take $\sim 19$ minutes to build its corresponding kernel on a $8$-vCPU Google Compute Engine (\texttt{c3-standard-8} GCE). This build time can be possibly reduced by running jobs on compute-heavy machines - reducing to $\sim 4$ to $5$ minutes on a powerful $88$-vCPU GCE (\texttt{c3-standard-88}).

\begin{table*}
\centering
\caption{Average (min/max) crashes resolved in $3$ runs with/without execution logs. Note that the SCS/LCS subsets of \data are those for which we could produce an execution trace, so they are not the same SCS/LCS subsets as in Table 2 of the paper.}
\begin{tabular}{c|l|c|c|c|c|c}
\toprule
\begin{tabular}[c]{@{}c@{}}Model\\ (Gemini)\end{tabular} &  \multicolumn{1}{c|}{Strategy} & $Traj$ & $\mathcal{D}$ & $\mathcal{B}$ & \begin{tabular}[c]{@{}c@{}}SCS Execution  \\ 111 bugs\end{tabular} & \begin{tabular}[c]{@{}c@{}}LCS Execution\\ 144 bugs\end{tabular} \\ \hline
1.5-pro-001 &  Hyp-SR then Patch-SC $\mathcal{F}$ & 4 & 4 & 1 & 54.67 & 56.33 \\ \hline
1.5-pro-001 &  Execution + Hyp-SR/Patch-SC $\mathcal{F}$ & 4 & 4 & 1 & \textbf{55} (51/57) & \textbf{64} (60/69) \\ \hline
1.5-pro-002 &  Hyp-SR then Patch-SC $\mathcal{F}$ & 4 & 4 & 1 & 58.33 & 64.33 \\ \hline
1.5-pro-002 &  Execution + Hyp-SR/Patch-SC $\mathcal{F}$ & 4 & 4 & 1 & \textbf{64} (62/68) & \textbf{71} (68/75) \\ \hline
2.5-pro &  Hyp-SR then Patch-SC $\mathcal{F}$ & 4 & 4 & 1 & 73.33 & 89 \\ \hline
2.5-pro &  Execution + Hyp-SR/Patch-SC $\mathcal{F}$ & 4 & 4 & 1 & \textbf{75} (72/79) & \textbf{102} (102/102) \\ 
\bottomrule
\end{tabular}
\label{tab:kbench_with_execution_results}
\end{table*}

We note that \sys adopts the above design as it serves to provide a generic building tool that can support numerous kernel versions. As each bug in \data addresses a different kernel version (git commit), the build configurations for each bug are also unique. Hence the current design serves to provide a uniform build process applicable to all the \data bugs ($279$ in total). However, to accurately mimic a crash-fixing workflow, the underlying platform must support ways to amortize the cost of compilation.

\textbf{\cache.} Thus, rather than starting compilation from scratch, \sysplus begins compilation from \cache - an ``intermediate cache'' unique to each sample in \kbenchsyz.  For each sample, \cache includes the source files and the pre-compiled object files for the entire Linux source code at the specified commit-id. Given that most human patches modify very few files, using \cache avoids rebuilding the kernel's \textit{unchanged} files; allowing for significant speedups (Linux has more than $50k$ files and $20M$ lines of code). An example of this speedup is shown in the last column of Figure \ref{fig:build-time-vs-core}. As seen, even when using an $88$-vCPU GCE (\texttt{c3-standard-88}), the time taken to build the kernel drops from $\sim 240$ to $135$ seconds. Additionally, when reducing the core count from $88$ to $8$, the ``from scratch'' build time increases $\sim 4.6$x, but the \cache build time increases by only  $\sim 1.2$x. Thus, for the purposes of ``crash resolution'', the performance of \sysplus is practically \textit{deployment agnostic}, achieving roughly the same performance despite downgrading from a $88$-vCPU to a $8$-vCPU GCE. 

\begin{figure}[htb]
    \centering
    \includegraphics[width=\linewidth]{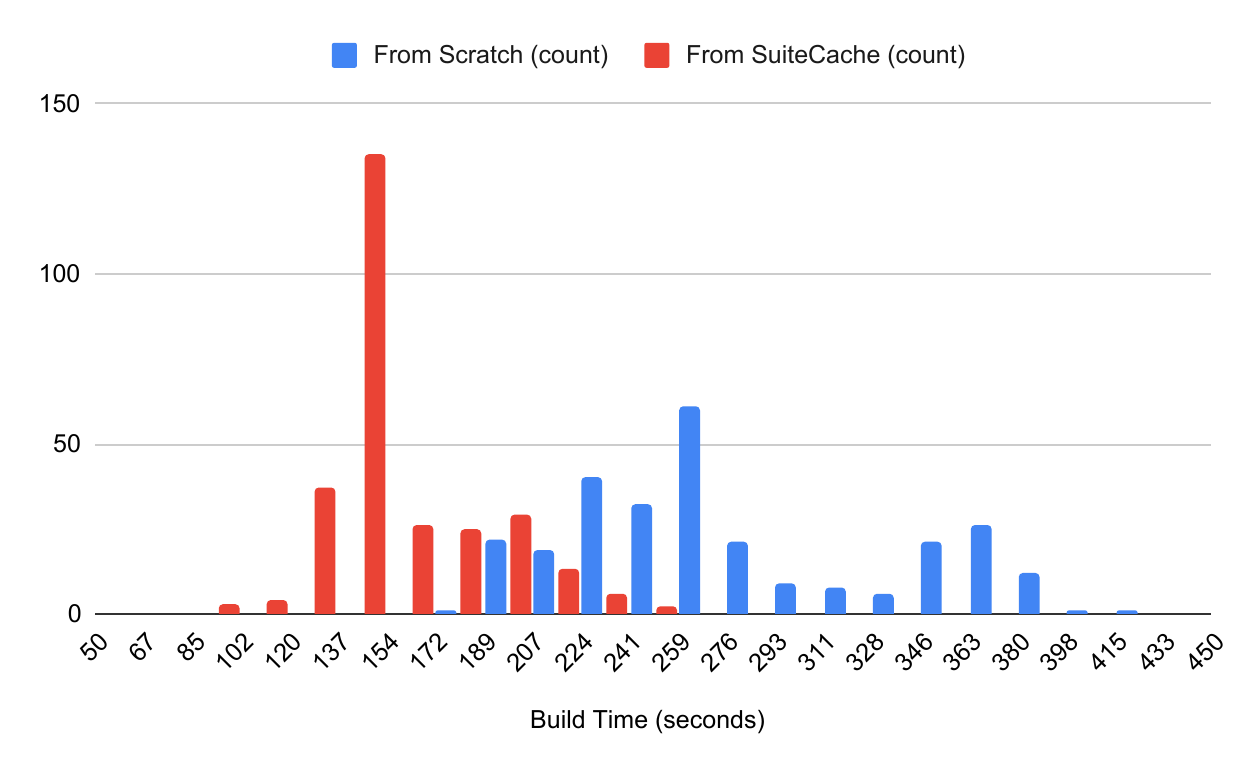}
    \caption{Distribution of kernel build time for \data}
    \label{fig:turnover-dist}
\end{figure}

To better understand the impact of using \cache, we plot the histogram distribution of the $88$-vCPU build times across the entire \data dataset in Figure \ref{fig:turnover-dist}.
As shown, the distribution of kernel build times in the entire \kbenchsyz dataset shifts significantly to the left when using \cache. On average, we measure a meaningful speedup of $\textbf{42.69\%}$ in kernel-build times using \cache.

\textbf{Compilation checks.} \sys's current design makes it impossible for users to issue ``compilation check'' jobs that specifically check if a patch is compilable. This tool proves useful if \agent needs to \textit{choose} a single patch from a collection of plausible patches \textit{before} attempting an entire kernel-build and reproducer run. Currently, the only way to perform this with \sys is to issue multiple kernel-build jobs and to wait until the jobs succeed or fail.

To enable quick compilation checks, we revisit the kernel compilation process. Importantly, we realize that in practice, to confirm if a patch is compilable, it seems sufficient to check if we can successfully compile the files that are modified by the patch. We validate this heuristic in our use of \sysplus by confirming that we do not observe linker failures after this pre-check. Thus, compiling other unmodified kernel files and linking objects into a kernel binary are, in practice, unnecessary steps that \sysplus can skip. Both these optimizations can be conveniently solved with the help of \cache. More specifically, when performing a compilation-check job, \sysplus pulls the relevant \cache into the working directory and can directly check if a patch is compilable by compiling solely the modified files and entirely skipping the linking process. This results in extremely quick turnaround times. In practice, \sysplus spends most of the time downloading and extracting \cache and $\sim 2$ to $3$ seconds per patch to check if the patch is compilable. As a result, \sysplus can conveniently check $10$ patches within a span of $30$ seconds on a $88$-vCPU machine.

Thus \cache is a simple yet efficient solution to both \textit{slow kernel builds} and \textit{slow compilation checks}.

\section{Evaluation Result with Execution Traces}
\label{sec:execution_table_results}

 In Table \ref{tab:kbench_with_execution_results}, we report our results when running \agent with execution information. For a fair comparison, we report numbers on the subset of data for which our execution collection process works ($255$/$279$ data points). This ``execution'' subset removes $6$ SCS bugs and $18$ LCS bugs. As shown, when using execution traces, we see relative improvements in both LCS and SCS. In LCS, we see relative improvements of $14.6\%$ for 2.5-pro, $10.36\%$ for 1.5-pro-002, and $13.61\%$ for 1.5-pro-001. In SCS, we see a $9.7\%$ relative increase for 1.5-pro-002 and consistent results for 1.5-pro-001 and 2.5-pro.

\section{Memory leak bug execution traces}
\label{sec:mem_leak_traces}

\begin{figure}[htb]
    \centering
    \includegraphics[width=0.8\linewidth]{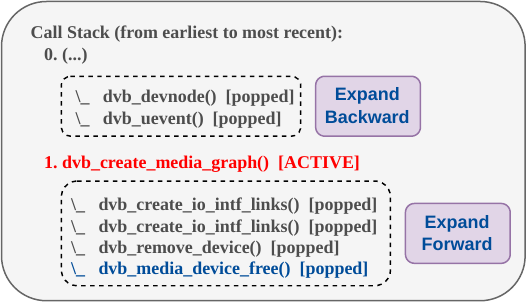}
    \caption{Minimized execution trace for the sample memory leak bug.}
    \label{fig:minimized_execution}
\end{figure}

In Figure \ref{fig:minimized_execution}, we depict the minimized execution traces for our example memory leak bug. As shown, ``dvb\_create\_media\_graph'' is the function (in red highlights) present in the crash stack trace. The output from ``stack trace anchoring'' matches this function. Then the forward and backward expansions collect $4$ and $2$ functions, respectively. In particular, the execution log's last function -- `dvb\_media\_device\_free()'' is the cause of the memory leak. Exposing such a minimized execution trace helps \agent concentrate on important and relevant executed functions.

\section{Plausible patches for \data bugs}
\label{sec:kbench_plausible_patches}

In Table \ref{tab:perfect_patches}, we show examples of plausible patches for sample \data bugs. As seen, the plausible patches span various bug types like null pointer dereferences, memory leak bugs, out-of-bound bugs and warning bugs.

\begin{table}[ht]
\caption{\agent generated plausible patches for different bugs in \data.}

\resizebox{0.95\linewidth}{!}{

\begin{tabular}{l}
\toprule

\begin{tabular}[c]{@{}l@{}}
\textbf{Patch for Null Pointer De-reference Bug \cite{plausible_nptr}} \\
\begin{minipage}{\linewidth}
\begin{minted}{c}
- pm_runtime_get(&hu->serdev->dev);
- pm_runtime_mark_last_busy(&hu->serdev->dev);
- pm_runtime_put_autosuspend(&hu->serdev->dev);
+ if (hu->serdev) { // Add null check
+   pm_runtime_get(&hu->serdev->dev);
+   pm_runtime_mark_last_busy(&hu->serdev->dev);
+   pm_runtime_put_autosuspend(&hu->serdev->dev);
+ }
\end{minted}
\end{minipage} \\
\end{tabular} \\

\begin{tabular}[c]{@{}l@{}}
\textbf{Patch for Memory Leak Bug \cite{plausible_memleak}} \\ 
\begin{minipage}{\linewidth}
\begin{minted}{c}
ret = dvb_usb_generic_rw(d, st->data, 1, 
                    st->data, 3, 0);
if (ret < 0) {
    deb_rc("cinergyt2_power_ctrl() Failed to 
    retrieve sleep state info\n");
+   /* Release the allocated frontend resources */
+   if (adap->fe_adap[0].fe) {
+         dvb_frontend_detach(adap->fe_adap[0].fe);
+         adap->fe_adap[0].fe = NULL;
+   }
\end{minted}
\end{minipage} \\
\end{tabular} \\

\begin{tabular}[c]{@{}l@{}}
\textbf{Patch for Out of Bounds Bug \cite{plausible_oob}} \\
\begin{minipage}{\linewidth}
\begin{minted}{c}
if (nls_disk) {
+     if (dstlen <= 0) // Check for available space
+       goto out;
      size = nls_disk->uni2char(ch, dst, dstlen);
      if (size < 0) {
      if (size == -ENAMETOOLONG)
@@ -130,6 +132,8 @@ void hfs_asc2mac(…)
      dst += size;
      dstlen -= size;
      } else {
+       if (dstlen <= 0) // Check for available space
+           goto out;
      *dst++ = ch > 0xff ? '?' : ch;
      dstlen--;
}
\end{minted}
\end{minipage} \\
\end{tabular} \\

\begin{tabular}[c]{@{}l@{}}
\textbf{Patch for Warning Bug \cite{plausible_warn}} \\
\begin{minipage}{\linewidth}
\begin{minted}{c}
  m->clusterofs = le16_to_cpu(di->di_clusterofs);
+ if (m->clusterofs >= 1<<vi->z_logical_clusterbits) 
+ {
+       erofs_err(inode->i_sb, "invalid 
clusterofs %u, exceeding cluster size",
+       m->clusterofs);
+       return -EFSCORRUPTED;
+ }
  m->pblk = le32_to_cpu(di->di_u.blkaddr);
  break;
\end{minted}
\end{minipage} \\
\end{tabular} \\

\bottomrule
\end{tabular}

}
\label{tab:perfect_patches}
\end{table}

\section{Hypothesis Generation Prompt}
\label{sec:hyp_gen_prompt}

{\scriptsize

\begin{verbatim}
<issue>
{CRASH TEXT}
</issue>

<code>
[start of file_1]
{file_1 text}
[end of file_1]
[start of file_2]
{file_2 text}
[end of file_2]
....
</code>

<execution>
Call Stack (from earliest to most recent):

{execution trace here}
...
</execution>

**Task**: Based on the given issue, code
snippet and **execution** please provide
a high-level reasoning for the root cause
of the issue followed by a natural language
explanation of a possible solution.

**Instructions**:
1. Write a natural language explanation of
a solution in the format `<solution> ...
</solution>`.
2. Replace `...` with your solution explanation
between the `<solution>` and `</solution>` tags.
3. The solution explanation must **NOT** contain
**large** code snippets. It must be primarily a
natural language explanation with optional
references to variables and short pieces of code. 

**Example**:

<code>
[start of dir_X/dir_Y/script.c]
int calc_sum(int max_num) {
    int sum = 0;
    for (int var=0; var<=max_num; var++) {
        printf("A dummy loop\n");
        sum -= var;
    }
    return sum;
}

int main() {
    int MAX = 10;
    int sum = calc_sum(MAX);
    sum -= 3;
    printf("Sum of numbers from 1 to 10 is %d
", sum-1);
}
[end of dir_X/dir_Y/script.c]
</code>

<execution>
main()  [ACTIVE]
	\_   calc_sum()  [executed and popped]
</execution>

**Answer of Example**:

<solution>
The snippet should calculate and print the sum
of numbers from 1 to 10. As shown in the execution
logs, the `main()` function calls a helper
function `calc_sum()` to perform the summation.
However, the variable `sum` is decremented in
the loop instead of incremented. To correct this
bug we should modify the code in the `for loop`
so that the variable `sum` is incremented correctly.
Additionally, in the `main()` function, there is
an erroneous subtraction (`sum -= 3`) and a faulty
printf statement. To correct this bug, we should
(a) remove the extraneous subtraction and (b)
correct the printf statement to print `sum`
instead of `sum-1`.
</solution>

******************


IMPORTANT NOTE: The generated hypothesis **must** be
related to the functions in the <execution> ...
</execution> tags as those functions **DEFINITELY**
contain the buggy code. So please first inspect the
functions in the <execution> ... </execution> tags
and then suggest a hypothesis that involves modifying
the code in those executed functions. Any hypothesis
that does **not** involve modifying the code in the
functions **mentioned** in the <execution> ...
</execution> tags will very likely be a **wrong**
hypothesis.
\end{verbatim}
}

\section{Patch Generation Prompt}
\label{sec:patch_gen_prompt}

{\scriptsize
\begin{verbatim}
<issue>
{CRASH TEXT}
</issue>

<code>
[start of file_1]
{file_1 text}
[end of file_1]
[start of file_2]
{file_2 text}
[end of file_2]
....
</code>

<execution>
Call Stack (from earliest to most recent):

{execution trace here}
...
</execution>

<solution>
{high-level reasoning prompted previously}
</solution>

**Task:** Generate a list of code modifications
based on the provided code snippets, the issue
and the natural language explanation of a
possible solution.

### Instructions:
1. **Import necessary libraries** as needed.
2. **Analyze the execution** to identify the
modifications required.
3. **Analyze the code** to identify the
modifications required.
4. **Copy and repeat** the natural language
explanation of a possible solution provided
in `<solution>...</solution>` 
5. **Format** the modification list as shown below:
    - Use `<reason>...</reason>` to describe
the reason for the modification.
    - Use `<file>...</file>` to specify the file path.
    - Use `<original>...</original>` to include
the original code snippet.
    - Use `<patched>...</patched>` to include
the modified code.
6. **Multiple modifications** may be provided
if necessary.

### Format for Modifications:

<solution>
...
</solution>

```
// Modification 1
<reason>
...
</reason>
<file>
...
</file>
<original>
...
</original>
<patched>
...
</patched>

// Modification 2
<reason>
...
</reason>
<file>
...
</file>
<original>
...
</original>
<patched>
...
</patched>

// Modification 3
...
```

### Example:

**Code Snippet:**

<code>
[start of dir_X/dir_Y/script.c]
int calc_sum(int max_num) {
    int sum = 0;
    for (int var=0; var<=max_num; var++) {
        printf("A dummy loop\n");
        sum -= var;
    }
    return sum;
}

int main() {
    int MAX = 10;
    int sum = calc_sum(MAX);
    sum -= 3;
    printf("Sum of numbers from 1 to 10 is %d
", sum-1);
}
[end of dir_X/dir_Y/script.c]
</code>

<execution>
main()  [ACTIVE]
	\_   calc_sum()  [executed and popped]
</execution>

<solution>
The snippet should calculate and print the
sum of numbers from 1 to 10. As shown in
the execution logs, the `main()` function
calls a helper function `calc_sum()` to
perform the summation. However, the variable
`sum` is decremented in the loop instead of
incremented. To correct this bug we should
modify the code in the `for loop` so that
the variable `sum` is incremented correctly.
Additionally, in the `main()` function, there
is an erroneous subtraction (`sum -= 3`) and
a faulty printf statement. To correct this
bug, we should (a) remove the extraneous
subtraction and (b) correct the printf
statement to print `sum` instead of `sum-1`.
</solution>

**Sample Answer:**

<solution>
The snippet should calculate and print the
sum of numbers from 1 to 10. As shown in
the execution logs, the `main()` function
calls a helper function `calc_sum()` to
perform the summation. However, the variable
`sum` is decremented in the loop instead of
incremented. To correct this bug we should
modify the code in the `for loop` so that the
variable `sum` is incremented correctly.
Additionally, in the `main()` function, there
is an erroneous subtraction (`sum -= 3`) and
a faulty printf statement. To correct this
bug, we should (a) remove the extraneous
subtraction and (b) correct the printf
statement to print `sum` instead of `sum-1`.
</solution>

```
// Modification 1
<reason>
The below change makes sure that the variable
"sum" is being incremented correctly inside
the while loop.
</reason>
<file>
dir_X/dir_Y/script.c
</file>
<original>
    for (int var=0; var<=max_num; var++) {
        printf("A dummy loop\n");
        sum -= var;
    }
</original>
<patched>
    for (int var=0; var<=max_num; var++) {
        printf("A dummy loop\n");
        sum += var; /* changed the
        subtraction to addition */
    }
</patched>

// Modification 2
<reason>
The below change makes sure that the extra
erroneous subtractions are removed and the
printf statement is corrected.
</reason>
<file>
dir_X/dir_Y/script.c
</file>
<original>
    sum -= 3;
    printf("Sum of numbers from 1 to 10 is %d
", sum-1);
</original>
<patched>
    printf("Sum of numbers from 1 to 10 is %d
", sum);
</patched>
```
******************
\end{verbatim}
}

\end{document}